\documentclass{PoS}

\usepackage[caption=false]{subfig}
\usepackage{graphicx}
\usepackage{bm}
\usepackage{amssymb}
\usepackage{amsmath}
\usepackage{amsfonts}
\usepackage[utf8]{inputenc}
\usepackage{placeins}
\usepackage{booktabs}
\usepackage{gensymb}
\usepackage{microtype}
\usepackage[utf8]{inputenc}
\usepackage{lineno}
\usepackage{multirow, makecell}

\usepackage{listings}
\usepackage{siunitx}
\usepackage{xspace}
\usepackage{floatrow}
\usepackage{comment}
\usepackage{resizegather}

\let\OLDthebibliography\thebibliography
\renewcommand\thebibliography[1]{
  \OLDthebibliography{#1}
  \setlength{\parskip}{0pt}
  \setlength{\itemsep}{0pt plus 0.3ex}
}

\DeclareSIUnit \s {\second}
\DeclareSIUnit \MB {\mega\byte}
\DeclareSIUnit \GB {\giga\byte}
\DeclareSIUnit \TB {\tera\byte}
\DeclareSIUnit \PB {\peta\byte}
\DeclareSIUnit \Mbps {\mega\bit/\s}
\DeclareSIUnit \Gbps {\giga\bit/\s}
\DeclareSIUnit \Tbps {\tera\bit/\s}
\DeclareSIUnit \Pbps {\peta\bit/\s}
\DeclareSIUnit \kton {\kilo\tonne} 
\DeclareSIUnit \kt {\kilo\tonne}
\DeclareSIUnit \Mt {\mega\tonne}
\DeclareSIUnit \eV {\electronvolt}
\DeclareSIUnit \keV {\kilo\electronvolt}
\DeclareSIUnit \MeV {\mega\electronvolt}
\DeclareSIUnit \GeV {\giga\electronvolt}
\DeclareSIUnit \TeV {\tera\electronvolt}
\DeclareSIUnit \PeV {\peta\electronvolt}
\DeclareSIUnit \EeV {\exa\electronvolt}
\DeclareSIUnit \m {\meter}
\DeclareSIUnit \cm {\centi\meter}
\DeclareSIUnit \in {\inchcommand}
\DeclareSIUnit \km {\kilo\meter}
\DeclareSIUnit \kV {\kilo\volt}
\DeclareSIUnit \kW {\kilo\watt}
\DeclareSIUnit \MW {\mega\watt}
\DeclareSIUnit \MHz {\mega\hertz}
\DeclareSIUnit \mrad {\milli\radian}
\DeclareSIUnit \sr {\steradian}
\DeclareSIUnit \year {years}
\DeclareSIUnit \day {days}
\DeclareSIUnit \POT {POT}
\DeclareSIUnit \sig {$\sigma$}
\DeclareSIUnit\parsec{pc}
\DeclareSIUnit\lightyear{ly}
\DeclareSIUnit\foot{ft}
\DeclareSIUnit\ft{ft}
\DeclareSIUnit \ppb{ppb}
\DeclareSIUnit \ppt{ppt}
\DeclareSIUnit \samples{S}
\DeclareSIUnit \pe{PE}

\newcommand\SigmaOne{\SI{68.3}\percent}
\newcommand\SigmaTwo{\SI{95.4}\percent}

\newcommand\LivetimeYears{\SI{7.5}\year\xspace}
\newcommand\LivetimeDays{\SI{2635}\day\xspace}

\newcommand\astronorm{\Phi_\texttt{astro}}
\newcommand\astrodeltagamma{\gamma_\texttt{astro}}
\newcommand\convnorm{\Phi_\texttt{conv}}
\newcommand\promptnorm{\Phi_\texttt{prompt}}
\newcommand\pik{R_{K/\pi}}
\newcommand\atmonunubar{{2\nu/\left(\nu+\bar{\nu}\right)}_\texttt{atmo}}
\newcommand\crdeltagamma{\Delta\gamma_\texttt{CR}}
\newcommand\muonnorm{\Phi_\mu}
\newcommand\domeff{\epsilon_\texttt{DOM}}
\newcommand\holeice{\epsilon_\texttt{head-on}}
\newcommand\anisotropy{a_s}

\usepackage{xparse}

\ExplSyntaxOn
\DeclareExpandableDocumentCommand{\eval}{m}{\fp_eval:n {#1}}
\ExplSyntaxOff

\newcommand\Stecker{Stecker~\cite{Stecker:2013fxa}}
\newcommand\SteckerBayes{4.32\times10^{-13}}
\newcommand\SteckerSPLBayes{1.45\times10^{-10}}
\newcommand\SteckerSPLPhiLower{2.95}
\newcommand\SteckerSPLPhiMode{4.08}
\newcommand\SteckerSPLPhiUpper{5.88}
\newcommand\SteckerSPLPhiLowerDelta{\eval{\SteckerSPLPhiMode - \SteckerSPLPhiLower}}
\newcommand\SteckerSPLPhiUpperDelta{\eval{\SteckerSPLPhiUpper - \SteckerSPLPhiMode}}
\newcommand\SteckerSPLPhiSummary{\SteckerSPLPhiMode^{+\SteckerSPLPhiUpperDelta}_{-\SteckerSPLPhiLowerDelta}}
\newcommand\SteckerSPLGammaLower{3.5}
\newcommand\SteckerSPLGammaMode{3.97}
\newcommand\SteckerSPLGammaUpper{4.51}
\newcommand\SteckerSPLGammaLowerDelta{\eval{\SteckerSPLGammaMode - \SteckerSPLGammaLower}}
\newcommand\SteckerSPLGammaUpperDelta{\eval{\SteckerSPLGammaUpper - \SteckerSPLGammaMode}}
\newcommand\SteckerSPLGammaSummary{\SteckerSPLGammaMode^{+\SteckerSPLGammaUpperDelta}_{-\SteckerSPLGammaLowerDelta}}
\newcommand\SteckerTableSummary{\Stecker & $\SteckerBayes$ & $\SteckerSPLBayes$ & $\SteckerSPLGammaSummary$ & $\SteckerSPLPhiSummary$}
\newcommand\Fang{Fang et al.~\cite{Fang:2017zjf}}
\newcommand\FangBayes{0.281}
\newcommand\FangSPLBayes{0.248}
\newcommand\FangSPLPhiLower{1.12}
\newcommand\FangSPLPhiMode{2.56}
\newcommand\FangSPLPhiUpper{3.84}
\newcommand\FangSPLPhiLowerDelta{\eval{\FangSPLPhiMode - \FangSPLPhiLower}}
\newcommand\FangSPLPhiUpperDelta{\eval{\FangSPLPhiUpper - \FangSPLPhiMode}}
\newcommand\FangSPLPhiSummary{\FangSPLPhiMode^{+\FangSPLPhiUpperDelta}_{-\FangSPLPhiLowerDelta}}
\newcommand\FangSPLGammaLower{3.33}
\newcommand\FangSPLGammaMode{3.83}
\newcommand\FangSPLGammaUpper{4.64}
\newcommand\FangSPLGammaLowerDelta{\eval{\FangSPLGammaMode - \FangSPLGammaLower}}
\newcommand\FangSPLGammaUpperDelta{\eval{\FangSPLGammaUpper - \FangSPLGammaMode}}
\newcommand\FangSPLGammaSummary{\FangSPLGammaMode^{+\FangSPLGammaUpperDelta}_{-\FangSPLGammaLowerDelta}}
\newcommand\FangTableSummary{\Fang & $\FangBayes$ & $\FangSPLBayes$ & $\FangSPLGammaSummary$ & $\FangSPLPhiSummary$}
\newcommand\KimuraBOne{Kimura et al. (B1)~\cite{Kimura:2014jba}}
\newcommand\KimuraBOneBayes{4.84\times10^{-6}}
\newcommand\KimuraBOneSPLBayes{8.38\times10^{-7}}
\newcommand\KimuraBOneSPLPhiLower{0.0}
\newcommand\KimuraBOneSPLPhiMode{0.98}
\newcommand\KimuraBOneSPLPhiUpper{2.02}
\newcommand\KimuraBOneSPLPhiLowerDelta{\eval{\KimuraBOneSPLPhiMode - \KimuraBOneSPLPhiLower}}
\newcommand\KimuraBOneSPLPhiUpperDelta{\eval{\KimuraBOneSPLPhiUpper - \KimuraBOneSPLPhiMode}}
\newcommand\KimuraBOneSPLPhiSummary{\KimuraBOneSPLPhiMode^{+\KimuraBOneSPLPhiUpperDelta}_{-\KimuraBOneSPLPhiLowerDelta}}
\newcommand\KimuraBOneSPLGammaLower{3.83}
\newcommand\KimuraBOneSPLGammaMode{4.5}
\newcommand\KimuraBOneSPLGammaUpper{5.0}
\newcommand\KimuraBOneSPLGammaLowerDelta{\eval{\KimuraBOneSPLGammaMode - \KimuraBOneSPLGammaLower}}
\newcommand\KimuraBOneSPLGammaUpperDelta{\eval{\KimuraBOneSPLGammaUpper - \KimuraBOneSPLGammaMode}}
\newcommand\KimuraBOneSPLGammaSummary{\KimuraBOneSPLGammaMode^{+\KimuraBOneSPLGammaUpperDelta}_{-\KimuraBOneSPLGammaLowerDelta}}
\newcommand\KimuraBOneTableSummary{\KimuraBOne & $\KimuraBOneBayes$ & $\KimuraBOneSPLBayes$ & $\KimuraBOneSPLGammaSummary$ & $\KimuraBOneSPLPhiSummary$}
\newcommand\KimuraBFour{Kimura et al. (B4)~\cite{Kimura:2014jba}}
\newcommand\KimuraBFourBayes{3.44\times10^{-4}}
\newcommand\KimuraBFourSPLBayes{0.666}
\newcommand\KimuraBFourSPLPhiLower{0.62}
\newcommand\KimuraBFourSPLPhiMode{1.39}
\newcommand\KimuraBFourSPLPhiUpper{2.57}
\newcommand\KimuraBFourSPLPhiLowerDelta{\eval{\KimuraBFourSPLPhiMode - \KimuraBFourSPLPhiLower}}
\newcommand\KimuraBFourSPLPhiUpperDelta{\eval{\KimuraBFourSPLPhiUpper - \KimuraBFourSPLPhiMode}}
\newcommand\KimuraBFourSPLPhiSummary{\KimuraBFourSPLPhiMode^{+\KimuraBFourSPLPhiUpperDelta}_{-\KimuraBFourSPLPhiLowerDelta}}
\newcommand\KimuraBFourSPLGammaLower{2.17}
\newcommand\KimuraBFourSPLGammaMode{2.43}
\newcommand\KimuraBFourSPLGammaUpper{2.74}
\newcommand\KimuraBFourSPLGammaLowerDelta{\eval{\KimuraBFourSPLGammaMode - \KimuraBFourSPLGammaLower}}
\newcommand\KimuraBFourSPLGammaUpperDelta{\eval{\KimuraBFourSPLGammaUpper - \KimuraBFourSPLGammaMode}}
\newcommand\KimuraBFourSPLGammaSummary{\KimuraBFourSPLGammaMode^{+\KimuraBFourSPLGammaUpperDelta}_{-\KimuraBFourSPLGammaLowerDelta}}
\newcommand\KimuraBFourTableSummary{\KimuraBFour & $\KimuraBFourBayes$ & $\KimuraBFourSPLBayes$ & $\KimuraBFourSPLGammaSummary$ & $\KimuraBFourSPLPhiSummary$}
\newcommand\KimuraTwoComp{Kimura et al. (two component)~\cite{Kimura:2014jba}}
\newcommand\KimuraTwoCompBayes{1.73\times10^{-4}}
\newcommand\KimuraTwoCompSPLBayes{6.12\times10^{-6}}
\newcommand\KimuraTwoCompSPLPhiLower{0.0}
\newcommand\KimuraTwoCompSPLPhiMode{0.0}
\newcommand\KimuraTwoCompSPLPhiUpper{0.69}
\newcommand\KimuraTwoCompSPLPhiLowerDelta{\eval{\KimuraTwoCompSPLPhiMode - \KimuraTwoCompSPLPhiLower}}
\newcommand\KimuraTwoCompSPLPhiUpperDelta{\eval{\KimuraTwoCompSPLPhiUpper - \KimuraTwoCompSPLPhiMode}}
\newcommand\KimuraTwoCompSPLPhiSummary{\KimuraTwoCompSPLPhiMode^{+\KimuraTwoCompSPLPhiUpperDelta}_{-\KimuraTwoCompSPLPhiLowerDelta}}
\newcommand\KimuraTwoCompSPLGammaLower{3.42}
\newcommand\KimuraTwoCompSPLGammaMode{4.15}
\newcommand\KimuraTwoCompSPLGammaUpper{4.99}
\newcommand\KimuraTwoCompSPLGammaLowerDelta{\eval{\KimuraTwoCompSPLGammaMode - \KimuraTwoCompSPLGammaLower}}
\newcommand\KimuraTwoCompSPLGammaUpperDelta{\eval{\KimuraTwoCompSPLGammaUpper - \KimuraTwoCompSPLGammaMode}}
\newcommand\KimuraTwoCompSPLGammaSummary{\KimuraTwoCompSPLGammaMode^{+\KimuraTwoCompSPLGammaUpperDelta}_{-\KimuraTwoCompSPLGammaLowerDelta}}
\newcommand\KimuraTwoCompTableSummary{\KimuraTwoComp & $\KimuraTwoCompBayes$ & $\KimuraTwoCompSPLBayes$ & $\KimuraTwoCompSPLGammaSummary$ & $\KimuraTwoCompSPLPhiSummary$}
\newcommand\MariaBLLacs{Padovani et al.~\cite{Padovani:2015mba}}
\newcommand\MariaBLLacsBayes{6.20\times10^{-11}}
\newcommand\MariaBLLacsSPLBayes{3.32\times10^{-7}}
\newcommand\MariaBLLacsSPLPhiLower{3.51}
\newcommand\MariaBLLacsSPLPhiMode{4.97}
\newcommand\MariaBLLacsSPLPhiUpper{6.65}
\newcommand\MariaBLLacsSPLPhiLowerDelta{\eval{\MariaBLLacsSPLPhiMode - \MariaBLLacsSPLPhiLower}}
\newcommand\MariaBLLacsSPLPhiUpperDelta{\eval{\MariaBLLacsSPLPhiUpper - \MariaBLLacsSPLPhiMode}}
\newcommand\MariaBLLacsSPLPhiSummary{\MariaBLLacsSPLPhiMode^{+\MariaBLLacsSPLPhiUpperDelta}_{-\MariaBLLacsSPLPhiLowerDelta}}
\newcommand\MariaBLLacsSPLGammaLower{3.25}
\newcommand\MariaBLLacsSPLGammaMode{3.59}
\newcommand\MariaBLLacsSPLGammaUpper{4.18}
\newcommand\MariaBLLacsSPLGammaLowerDelta{\eval{\MariaBLLacsSPLGammaMode - \MariaBLLacsSPLGammaLower}}
\newcommand\MariaBLLacsSPLGammaUpperDelta{\eval{\MariaBLLacsSPLGammaUpper - \MariaBLLacsSPLGammaMode}}
\newcommand\MariaBLLacsSPLGammaSummary{\MariaBLLacsSPLGammaMode^{+\MariaBLLacsSPLGammaUpperDelta}_{-\MariaBLLacsSPLGammaLowerDelta}}
\newcommand\MariaBLLacsTableSummary{\MariaBLLacs & $\MariaBLLacsBayes$ & $\MariaBLLacsSPLBayes$ & $\MariaBLLacsSPLGammaSummary$ & $\MariaBLLacsSPLPhiSummary$}
\newcommand\MurasechockedJets{Senno et al.~\cite{Senno:2015tsn}}
\newcommand\MurasechockedJetsBayes{0.256}
\newcommand\MurasechockedJetsSPLBayes{3.52}
\newcommand\MurasechockedJetsSPLPhiLower{2.02}
\newcommand\MurasechockedJetsSPLPhiMode{3.36}
\newcommand\MurasechockedJetsSPLPhiUpper{4.92}
\newcommand\MurasechockedJetsSPLPhiLowerDelta{\eval{\MurasechockedJetsSPLPhiMode - \MurasechockedJetsSPLPhiLower}}
\newcommand\MurasechockedJetsSPLPhiUpperDelta{\eval{\MurasechockedJetsSPLPhiUpper - \MurasechockedJetsSPLPhiMode}}
\newcommand\MurasechockedJetsSPLPhiSummary{\MurasechockedJetsSPLPhiMode^{+\MurasechockedJetsSPLPhiUpperDelta}_{-\MurasechockedJetsSPLPhiLowerDelta}}
\newcommand\MurasechockedJetsSPLGammaLower{3.05}
\newcommand\MurasechockedJetsSPLGammaMode{3.67}
\newcommand\MurasechockedJetsSPLGammaUpper{4.24}
\newcommand\MurasechockedJetsSPLGammaLowerDelta{\eval{\MurasechockedJetsSPLGammaMode - \MurasechockedJetsSPLGammaLower}}
\newcommand\MurasechockedJetsSPLGammaUpperDelta{\eval{\MurasechockedJetsSPLGammaUpper - \MurasechockedJetsSPLGammaMode}}
\newcommand\MurasechockedJetsSPLGammaSummary{\MurasechockedJetsSPLGammaMode^{+\MurasechockedJetsSPLGammaUpperDelta}_{-\MurasechockedJetsSPLGammaLowerDelta}}
\newcommand\MurasechockedJetsTableSummary{\MurasechockedJets & $\MurasechockedJetsBayes$ & $\MurasechockedJetsSPLBayes$ & $\MurasechockedJetsSPLGammaSummary$ & $\MurasechockedJetsSPLPhiSummary$}
\newcommand\SBGminBmodel{Bartos et al.~\cite{Bartos:2015xpa}}
\newcommand\SBGminBmodelBayes{1.15\times10^{-14}}
\newcommand\SBGminBmodelSPLBayes{2.81\times10^{-16}}
\newcommand\SBGminBmodelSPLPhiLower{0.0}
\newcommand\SBGminBmodelSPLPhiMode{0.0}
\newcommand\SBGminBmodelSPLPhiUpper{0.49}
\newcommand\SBGminBmodelSPLPhiLowerDelta{\eval{\SBGminBmodelSPLPhiMode - \SBGminBmodelSPLPhiLower}}
\newcommand\SBGminBmodelSPLPhiUpperDelta{\eval{\SBGminBmodelSPLPhiUpper - \SBGminBmodelSPLPhiMode}}
\newcommand\SBGminBmodelSPLPhiSummary{\SBGminBmodelSPLPhiMode^{+\SBGminBmodelSPLPhiUpperDelta}_{-\SBGminBmodelSPLPhiLowerDelta}}
\newcommand\SBGminBmodelSPLGammaLower{3.42}
\newcommand\SBGminBmodelSPLGammaMode{4.25}
\newcommand\SBGminBmodelSPLGammaUpper{5.0}
\newcommand\SBGminBmodelSPLGammaLowerDelta{\eval{\SBGminBmodelSPLGammaMode - \SBGminBmodelSPLGammaLower}}
\newcommand\SBGminBmodelSPLGammaUpperDelta{\eval{\SBGminBmodelSPLGammaUpper - \SBGminBmodelSPLGammaMode}}
\newcommand\SBGminBmodelSPLGammaSummary{\SBGminBmodelSPLGammaMode^{+\SBGminBmodelSPLGammaUpperDelta}_{-\SBGminBmodelSPLGammaLowerDelta}}
\newcommand\SBGminBmodelTableSummary{\SBGminBmodel & $\SBGminBmodelBayes$ & $\SBGminBmodelSPLBayes$ & $\SBGminBmodelSPLGammaSummary$ & $\SBGminBmodelSPLPhiSummary$}
\newcommand\TavecchilowPower{Tavecchio et al.~\cite{Tavecchio:2014eia}}
\newcommand\TavecchilowPowerBayes{0.0730}
\newcommand\TavecchilowPowerSPLBayes{1.04}
\newcommand\TavecchilowPowerSPLPhiLower{2.22}
\newcommand\TavecchilowPowerSPLPhiMode{3.7}
\newcommand\TavecchilowPowerSPLPhiUpper{5.09}
\newcommand\TavecchilowPowerSPLPhiLowerDelta{\eval{\TavecchilowPowerSPLPhiMode - \TavecchilowPowerSPLPhiLower}}
\newcommand\TavecchilowPowerSPLPhiUpperDelta{\eval{\TavecchilowPowerSPLPhiUpper - \TavecchilowPowerSPLPhiMode}}
\newcommand\TavecchilowPowerSPLPhiSummary{\TavecchilowPowerSPLPhiMode^{+\TavecchilowPowerSPLPhiUpperDelta}_{-\TavecchilowPowerSPLPhiLowerDelta}}
\newcommand\TavecchilowPowerSPLGammaLower{3.39}
\newcommand\TavecchilowPowerSPLGammaMode{3.88}
\newcommand\TavecchilowPowerSPLGammaUpper{4.53}
\newcommand\TavecchilowPowerSPLGammaLowerDelta{\eval{\TavecchilowPowerSPLGammaMode - \TavecchilowPowerSPLGammaLower}}
\newcommand\TavecchilowPowerSPLGammaUpperDelta{\eval{\TavecchilowPowerSPLGammaUpper - \TavecchilowPowerSPLGammaMode}}
\newcommand\TavecchilowPowerSPLGammaSummary{\TavecchilowPowerSPLGammaMode^{+\TavecchilowPowerSPLGammaUpperDelta}_{-\TavecchilowPowerSPLGammaLowerDelta}}
\newcommand\TavecchilowPowerTableSummary{\TavecchilowPower & $\TavecchilowPowerBayes$ & $\TavecchilowPowerSPLBayes$ & $\TavecchilowPowerSPLGammaSummary$ & $\TavecchilowPowerSPLPhiSummary$}
\newcommand\TDEWinterBiehl{Biehl et al.~\cite{Biehl:2017hnb}}
\newcommand\TDEWinterBiehlBayes{8.66\times10^{-7}}
\newcommand\TDEWinterBiehlSPLBayes{0.362}
\newcommand\TDEWinterBiehlSPLPhiLower{4.06}
\newcommand\TDEWinterBiehlSPLPhiMode{5.09}
\newcommand\TDEWinterBiehlSPLPhiUpper{7.16}
\newcommand\TDEWinterBiehlSPLPhiLowerDelta{\eval{\TDEWinterBiehlSPLPhiMode - \TDEWinterBiehlSPLPhiLower}}
\newcommand\TDEWinterBiehlSPLPhiUpperDelta{\eval{\TDEWinterBiehlSPLPhiUpper - \TDEWinterBiehlSPLPhiMode}}
\newcommand\TDEWinterBiehlSPLPhiSummary{\TDEWinterBiehlSPLPhiMode^{+\TDEWinterBiehlSPLPhiUpperDelta}_{-\TDEWinterBiehlSPLPhiLowerDelta}}
\newcommand\TDEWinterBiehlSPLGammaLower{2.97}
\newcommand\TDEWinterBiehlSPLGammaMode{3.35}
\newcommand\TDEWinterBiehlSPLGammaUpper{3.75}
\newcommand\TDEWinterBiehlSPLGammaLowerDelta{\eval{\TDEWinterBiehlSPLGammaMode - \TDEWinterBiehlSPLGammaLower}}
\newcommand\TDEWinterBiehlSPLGammaUpperDelta{\eval{\TDEWinterBiehlSPLGammaUpper - \TDEWinterBiehlSPLGammaMode}}
\newcommand\TDEWinterBiehlSPLGammaSummary{\TDEWinterBiehlSPLGammaMode^{+\TDEWinterBiehlSPLGammaUpperDelta}_{-\TDEWinterBiehlSPLGammaLowerDelta}}
\newcommand\TDEWinterBiehlTableSummary{\TDEWinterBiehl & $\TDEWinterBiehlBayes$ & $\TDEWinterBiehlSPLBayes$ & $\TDEWinterBiehlSPLGammaSummary$ & $\TDEWinterBiehlSPLPhiSummary$}

\newcommand\SPLFreqBFNorm{6.45} 
\newcommand\SPLFreqWilksUpperNorm{7.91} 
\newcommand\SPLFreqWilksLowerNorm{5.99} 
\newcommand\SPLFreqWilksUpperNormDelta{\eval{\SPLFreqWilksUpperNorm-\SPLFreqBFNorm}} 
\newcommand\SPLFreqWilksLowerNormDelta{\eval{\SPLFreqBFNorm-\SPLFreqWilksLowerNorm}} 
\newcommand\SPLFreqWilksNormSummary{\SPLFreqBFNorm^{+\SPLFreqWilksUpperNormDelta}_{-\SPLFreqWilksLowerNormDelta}}

\newcommand\SPLFreqBFIndex{2.89} 
\newcommand\SPLFreqWilksUpperIndex{3.09} 
\newcommand\SPLFreqWilksLowerIndex{2.70} 
\newcommand\SPLFreqWilksUpperIndexDelta{\eval{\SPLFreqWilksUpperIndex-\SPLFreqBFIndex}} 
\newcommand\SPLFreqWilksLowerIndexDelta{\eval{\SPLFreqBFIndex-\SPLFreqWilksLowerIndex}} 
\newcommand\SPLFreqWilksIndexSummary{\SPLFreqBFIndex^{+\SPLFreqWilksUpperIndexDelta}_{-\SPLFreqWilksLowerIndexDelta}}

\title{Characterization of the Astrophysical Diffuse Neutrino Flux with IceCube High-Energy Starting Events}

\ShortTitle{HESE 7.5 Diffuse}

\author{
The IceCube Collaboration\footnote{For collaboration list, see PoS(ICRC2019) 1177.}\\
{\itshape \href{http://icecube.wisc.edu/collaboration/authors/icrc19_icecube}{http://icecube.wisc.edu/collaboration/authors/icrc19\_icecube}}\\
E-mail: \email{austin.schneider@icecube.wisc.edu}
}

\abstract{
The IceCube neutrino observatory has established the existence of an astrophysical diffuse neutrino component above $\sim\SI{100}\TeV$. This discovery was made using the high-energy starting event sample, which uses the outer layer of instrumented volume as a veto to significantly reduce atmospheric background. We present the latest astrophysical neutrino flux measurement using high-energy starting events. This latest iteration of the analysis extends the sample by $1.5$ years for a total of $\LivetimeYears$, updates the event properties with newer models of light transport in the glacial ice, and has an improved systematic treatment. As part of this new analysis, we report on compatibility of our observations with detailed isotropic flux models proposed in the literature as well as the standard generic models such as single, double power-law scenarios. We find that none of the tested models are substantially preferred with respect to a single power law.\\
\vspace{4mm}
{\bfseries Corresponding authors:}
\speaker{Austin Schneider}$^{1}$\\
{$^{1}$ \itshape Dept. of Physics and Wisconsin IceCube Particle Astrophysics Center, University of Wisconsin, Madison WI 53706, USA}
}

\FullConference{36th International Cosmic Ray Conference -ICRC2019-\\
		July 24th - August 1st, 2019\\
		Madison, WI, U.S.A.}

\begin{document}

\section{Introduction}
It has been more than one hundred years since Victor Hess discovered cosmic rays.
Since then cosmic rays have been carefully characterized by studying the showers they produce when interacting in the Earth's atmosphere and also by direct measurements in outer space.
Even though much progress has been made in measuring their energy distribution, composition, and arrival distribution, their sources remain a mystery.
This is mainly due to the fact that galactic and extragalactic magnetic fields deflect cosmic rays in transit obscuring their origin.
The most promising way of discovering the sources of the cosmic rays relies on observing other cosmic messengers that are not deflected by magnetic fields such as gamma-rays and neutrinos.

Neutrinos play a unique role in this puzzle as they not only point to their point of production, but are far less likely to be absorbed in transit, which is in contrast to gamma-rays which have a galactic scale mean free path at \si{PeV} energies.
Large-scale neutrino detectors like the IceCube South Pole Neutrino Observatory are capable of observing neutrinos in the \SI{10}\TeV{} to \SI{10}\PeV{} energy range where a signal of astrophysical neutrinos is discernable from atmospheric background.
We present an update to the selection and analysis of IceCube's High Energy Starting Events (HESE), a sample of high energy neutrinos with interaction vertices contained within the detector fiducial volume.
Previously this sample was analyzed with two~\cite{Aartsen:2013jdh}, three~\cite{Aartsen:2014gkd}, four~\cite{Aartsen:2015zva}, and six~\cite{Kopper:2017zzm} years of data, leading to the discovery of a high-energy astrophysical neutrino flux.
This work extends the sample with an additional one and a half years of data for a total of \LivetimeDays of livetime, but primarily improves the description of atmospheric background and the treatment of uncertainties.
The analysis is performed using binned likelihood based statistical techniques, where cascades, tracks, and double-cascades are binned separately.
There are 10 bins in $\cos(\theta_z)$ and 20 bins in $\log(E_\textmd{deposited})$ for tracks and cascades for a total of 200 bins per morphology.
Double-cascades are split into 10 bins in $\log(L_\textmd{reco})$ and 20 bins in $\log(E_\textmd{deposited})$ for a total of 200 bins.
Where $\theta_z$ is the reconstructed zenith angle, $E_\textmd{deposited}$ is the reconstructed deposited energy, and $L_\texttt{reco}$ is the reconstructed distance between energy depositions.

\section{Analysis Improvements}

\subsection{Event Categorization}

Most often IceCube events are categorized into two morphologies -- cascades and tracks -- which provide a handle on the astrophysical neutrino flavor composition\cite{Aartsen:2015ivb, Aartsen:2015knd,Vincent:2016nut}, but this morphological classification misses a third morphology accessible above \SI{10}\TeV{} produced by charged-current $\nu_\tau$ interactions~\cite{Cowen:2007ny}.
This work now includes a dedicated search for tau neutrino double-bang events, described in~\cite{Stachurska:2019icrc_tau}, which classifies events into three morphologies: tracks, cascades, and double-cascades.
This additional information provides a better handle on the flavor content of the flux.
Inline with the addition of this new event category, we also include an additional observable: the reconstructed distance between the cascades in a double-cascade event.

\subsection{Background Modelling}
Previous analysis iterations only allowed the normalizations of various components and the spectral index of the astrophysical components to vary.
These were: $\promptnorm$ for the normalization of the flux of atmospheric neutrinos produced by charmed hadrons, $\convnorm$ for the normalization of those produced by pions and kaons called the ``conventional flux'', and $\muonnorm$ for the normalization of penetrating atmospheric muons in the sample~\cite{Bhattacharya:2015jpa}.
The first improvement implemented in this work is the addition of model parameters that modify the atmospheric model templates used in the analysis.
We introduce three parameters: $\pik$, $\atmonunubar$, and $\crdeltagamma$.
$\pik$ alters the relative contribution of pions and kaons to the atmospheric neutrino flux, providing an effective treatment of the atmospheric production uncertainties in the energy range we are concerned with.
$\atmonunubar$ alters the relative contribution of neutrinos and anti-neutrinos from the atmospheric flux.
Finally, $\crdeltagamma$ modifies the spectrum of the atmospheric neutrino component to account for uncertainties in the cosmic ray spectrum.
These extensions of the background model account for much of the background modelling uncertainty.
Additionally, this work includes an update of the coincident penetrating muon atmospheric neutrino veto probability calculation~\cite{Arguelles:2018awr}.
In this sample the primary effect of the update to the veto calculation is a reduction of the expected number of atmospheric events in the down-going region.
This reduction is small for the atmospheric neutrino flux from pions and kaons, and larger for contributions that come from the prompt decay of charmed hadrons~\cite{Arguelles:2018awr}.
The contribution from neutrinos produced by charmed hadrons is small relative to the others, and so this improvement does not significantly change the fit of the astrophysical parameters~\cite{Bhattacharya:2015jpa}.

\subsection{Detector Systematics}
Aside from updates to the atmospheric neutrino flux modelling, we also account for the most relevant detector response uncertainties.
Through calibration efforts, IceCube has been able to measure the absolute efficiency of its optical modules~\cite{Aartsen:2016nxy}; however uncertainty in this measurement translates into non-negligible uncertainty in the energy scale.
To account for this, we introduce an efficiency parameter ($\domeff$) that modifies the photon detection efficiency relative to the baseline of each optical module.
In a similar vein, there is uncertainty in the angular acceptance of the modules that depends on the properties of the local ice surrounding them.
We introduce another parameter to the model, $\holeice$, which varies the head-on angular acceptance of the modules; see~\cite{Aartsen:2016nxy, Aartsen:2014yll, Aartsen:2017nmd} for details.
The last detector systematic included is the strength of the ice anisotropy, $\anisotropy$, which primarily affects the reconstructed length of double cascades as a function of their angle with respect to the anisotropy axis~\cite{Chirkin:2013}.
The effect of these parameters on the expected event rate is then computed by producing variants of the simulation for a discrete set of these parameters as in the case of $\domeff$ and $\holeice$, or by modifying the assumptions in the reconstruction as in the case of $\anisotropy$.
The relative change in event rate associated with these parameters is approximated by a b-spline fit~\cite{Weaver:2015thesis,Whitehorn:2013nh} to the simulation changes with respect to the nominal simulation and reconstruction.
This continuous parameterization is then used to reweight the nominal simulation.
In doing this we ignore correlations between these systematic effects~\cite{Aartsen:2018afa}.

\subsection{Simulation Statistical Uncertainty}
Statistical uncertainties from the limited simulation sample size are non-negligible.
Relative statistical uncertainty of the various neutrino component expectations are small.
However, atmospheric muon simulation is of small size and dominates the statistical uncertainty in some bins.
In such bins the statistical errors from data and simulation are comparable.
To account for this, we use a likelihood based method that incorporates simulation statistical uncertainties~\cite{SAYPaper}.
This method ensures that the uncertainties in the final analysis results account for statistical deficiencies.

\section{Characterizing the Diffuse Flux}
To date, the precise origin and production mechanism of astrophysical neutrinos remains unknown.
As a simplifying assumption in this work we assume that the incident astrophysical neutrino flux is isotropic, equal between neutrino flavors, and equal between neutrinos and anti-neutrinos.
These assumptions are reasonable given that, due to the finite energy resolution of the experiment and unknown source distance, neutrino oscillations are expected to yield very close to equal amounts of each flavor~\cite{Palladino:2015zua,Arguelles:2015dca}.
Finally, the isotropy is justified as, at the moment, there is no observational evidence for anisotropy in the high-energy neutrino sky~\cite{Aartsen:2018ywr,Aartsen:2018ppz}.
In light of this, we test a variety of assumptions for the astrophysical neutrino energy spectrum; the results of which are summarized in Tbl.~\ref{tbl:diffuse_models}.
We split the discussion of these results into three parts depending on the assumed spectral model: the benchmark single power-law model, other generic models, and specific models from the literature.

For frequentist results we will use the maximum likelihood estimator, of the profile likelihood, and Wilks' theorem to quote confidence regions.
In the model comparison table we report the Bayes factor for each scenario we consider.
The Bayes factor is defined as the ratio between the evidence of the alternative and null hypotheses, where the evidence is the integral of the posterior distribution with respect to all model parameters. Namely
\begin{equation}
	\mathcal{B} = \frac{\int d\vec\eta~\mathcal{L}_\texttt{\tiny{Alt}}(\vec\eta)}{\int d\vec\eta d\vec\Phi~\mathcal{L}_\texttt{\tiny{Null}}(\vec\Phi,\vec\eta)},
\end{equation}
where $\vec\eta$ are the nuisance parameters, and $\vec\Phi$ are the two parameters of the single power law.
These Bayes factors can be understood as follows.
If one assumes non-committal priors, the rejection of the null hypothesis with respect to the alternative corresponds to an odds of one in $\mathcal{B}$.
In all cases, the null hypothesis is the single power-law model.
We consider alternative scenarios that include only the specified model on its own, and scenarios where the specified model is accompanied by a power-law component.
For scenarios that are accompanied by a power-law component, we also report the maximum {\it a posteriori}\ (MAP) estimator of the power-law parameters and the $\SigmaOne$ highest posterior density regions (HPD).

\subsection{Single power law}
The single power-law model has two free parameters: normalization of the flux ($\Phi_\texttt{astro}$) and spectral index ($\gamma_\texttt{astro}$).
For this model the energy spectrum is
\begin{linenomath*}
    \begin{equation}
        \frac{d\Phi_{6\nu}}{dE}=\Phi_\texttt{astro}{\left(\frac{E_\nu}{100 \textmd{TeV}}\right)}^{-\gamma_\texttt{astro}} \cdot 10^{-18}~[\textmd{GeV}^{-1}\textmd{cm}^{-2}\textmd{s}^{-1}\textmd{sr}^{-1}],
        \label{eq:single_power_law_flux}
    \end{equation}
\end{linenomath*}
where \SI{100}\TeV{} is an arbitrary pivot point chosen to allow comparison with previous results, and both $\Phi_\texttt{astro}$ and $\Phi_{6\nu}$ are in terms of the flux summed over all neutrino species.

\begin{figure}
    \centering
    \includegraphics[width=0.49\linewidth]{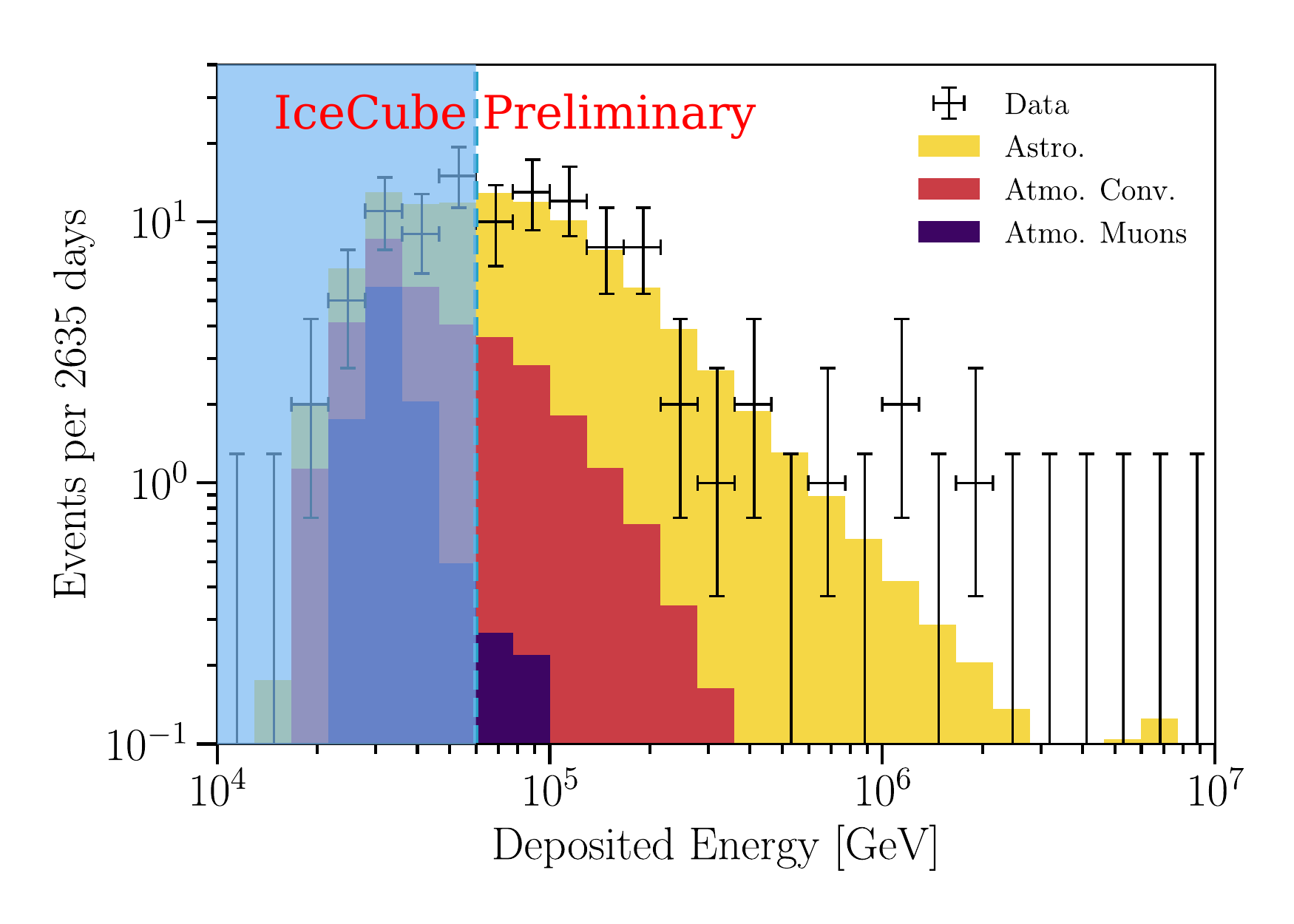}
    \includegraphics[width=0.49\linewidth]{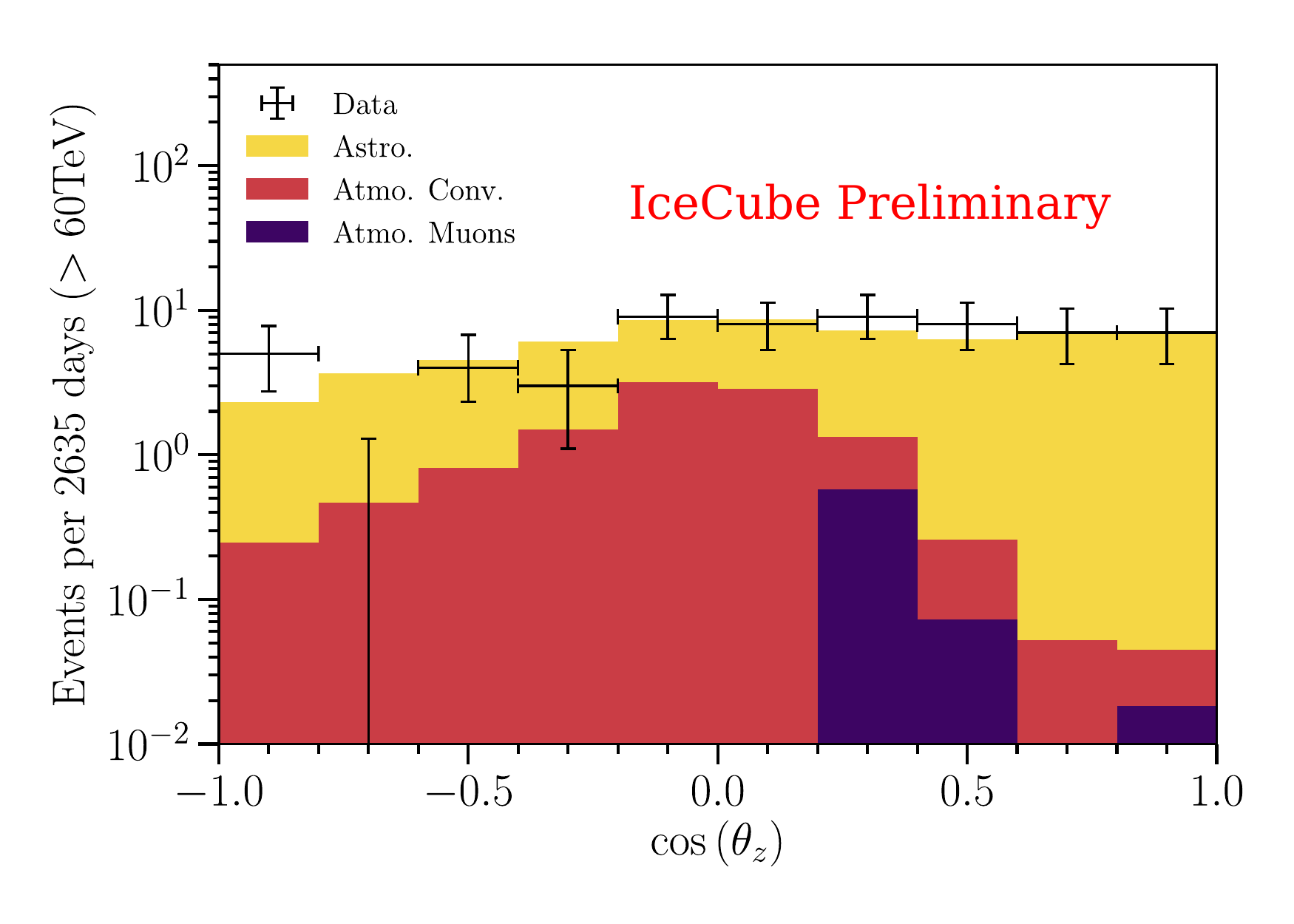}
    \caption{\textbf{\textit{Deposited energy and reconstructed $\cos\theta$ distributions.}} Data are shown as black crosses and the best-fit expectation as a stacked histogram with each color specifying a flux component: astrophysical (golden), conventional atmospheric as mentioned above (red), and penetrating muons (purple); the best fit prompt normalization is zero and is not shown. Left: distributions of observed and expected events as a function of the reconstructed deposited energy. Events below 60 TeV (light blue vertical line) are not included in the fit, but one sees good data-MC agreement extending into this energy range. Right: distribution of observed and expected best-fit events as a function of the cosine of the reconstructed zenith angle.}\label{fig:energy-zenith}
\end{figure}

We obtain a best-fit astrophysical spectral index ($\astrodeltagamma$) of $\SPLFreqWilksIndexSummary$ and a best fit astrophysical normalization ($\astronorm$) of $\SPLFreqWilksNormSummary$ where the errors correspond to the $\SigmaOne$ confidence interval in the aforementioned frequentist construction.
These best-fit parameters are compatible with the previous analysis using six~\cite{Kopper:2017zzm} years of data.
Figure~\ref{fig:energy-zenith} shows the data compared to the expected number of events assuming the best-fit parameters, as a function of the reconstructed deposited energy (left) and the cosine of the reconstructed zenith angle (right).
The relatively flat distribution in the cosine of the zenith angle cannot be reproduced by the atmospheric background components alone, leading to the high significance of the astrophysical component reported in previous analyses with respect to an atmospheric background only hypothesis.

\begin{figure}
\vspace{-4mm}
\floatbox[{\capbeside\thisfloatsetup{capbesideposition={right,center},capbesidewidth=0.5\textwidth}}]{figure}[\FBwidth]
{\caption{\textbf{\textit{Best fit parameters for the single power law.}} Contours in blue represent results from this work, while the orange contours show results from IceCube’s 9.5yr diffuse numu sample~\cite{Stettner:2019icrc_numu}, and the purple contours show results from IceCube’s multi-year cascade sample~\cite{Haack:2017dxi}. Solid contours represent the $\SigmaOne$ confidence regions, and dashed contours the $\SigmaTwo$ confidence regions.}\label{fig:SPL_frequentist}}
{\includegraphics[width=0.4\textwidth]{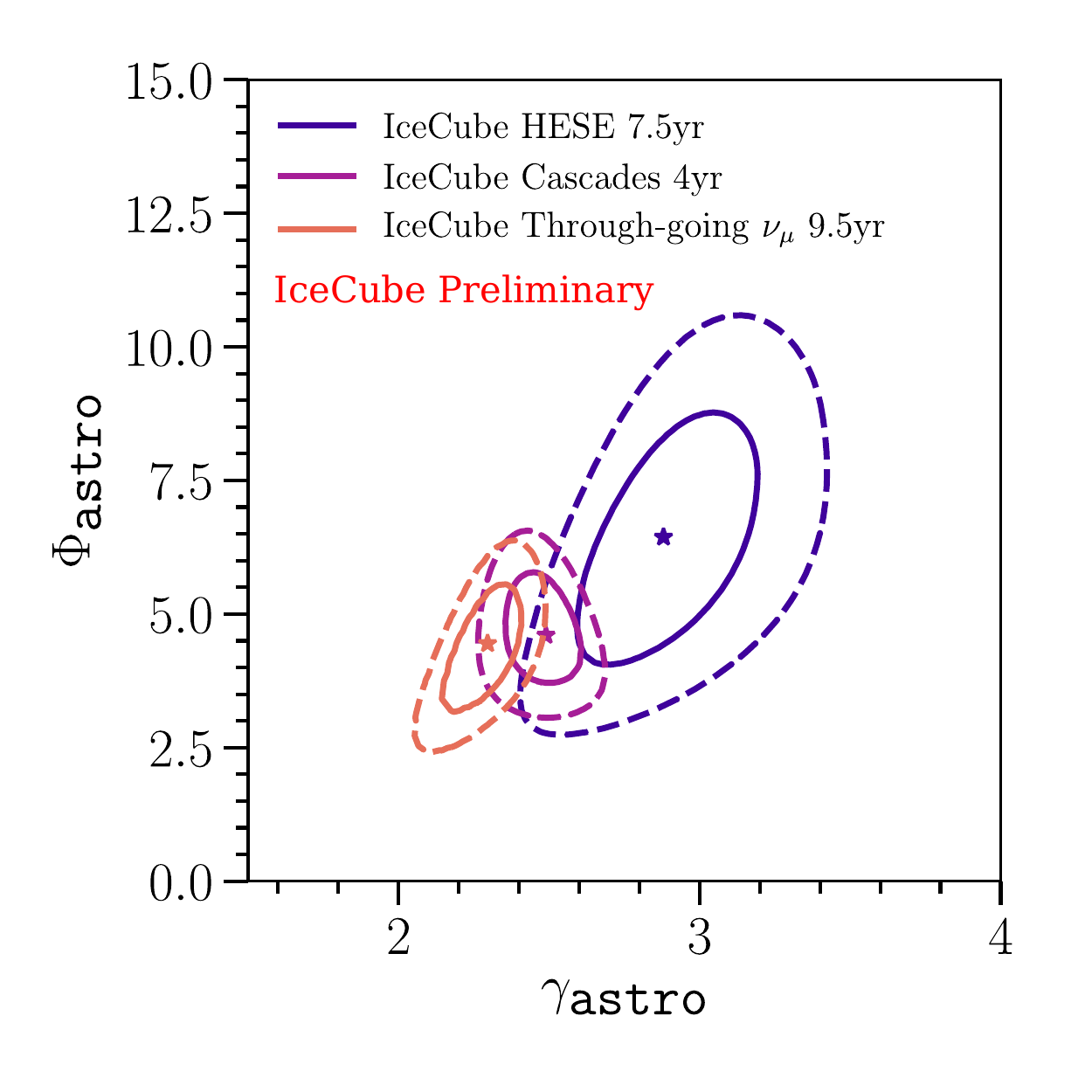}}
\vspace{-6mm}
\end{figure}

In figure~\ref{fig:SPL_frequentist}, the confidence regions for three different IceCube data-sets assuming an unbroken single power law are shown.
The three measurements have distinctly different best-fit points, but appear to be compatible with each other within their $\SigmaTwo$ regions.
In light of this we would like to consider explanations for the different spectral indices beyond pure statistical variations.
However, as the three data-sets make distinctly different energy cuts, select for different morphological classes of events, and have very different signal to background ratios in the energy ranges common between them, it is clear that they may be measuring different portions of the neutrino flux. A possible explanation for the different measurements may then be the existence of additional spectral structure. We consider other models beyond the na\"ive unbroken power-law assumption on these grounds.

\subsection{Other Generic Models}
We consider the possibility of two distinct astrophysical components that are both distributed as power laws in energy, and refer to this as the double power-law model.
The differential energy spectrum is then given by
\begin{linenomath*}
    \begin{gather}
    \begin{split}
        &\frac{d\Phi_{6\nu}}{dE} = \left[\Phi_\texttt{astro1}\left({\frac{E_\nu}{100 \textmd{TeV}}}\right)^{-\gamma_\texttt{astro1}} + \Phi_\texttt{astro2}{\left(\frac{E_\nu}{100 \textmd{TeV}}\right)}^{-\gamma_\texttt{astro2}}\right] \cdot 10^{-18}~[\textmd{GeV}^{-1}\textmd{cm}^{-2}\textmd{s}^{-1}\textmd{sr}^{-1}].
    \end{split}
    \label{eq:dpl_flux}
    \end{gather}
\end{linenomath*}
This model has two regions of the phase space in which it is equivalent to the single power-law model.
In tests of this model the $\SigmaOne$ HPD regions in the relevant two-dimensional projections of the posterior distribution contain the regions of phase space equivalent to the single power law.
On this basis, we conclude that there is not substantial power in this sample to discriminate between the double power-law and single power-law models.

\subsection{Specific Models}
In this work we also compare the observation to models published in the literature.
These models consider the following sources categories: AGN, low-luminosity AGN BLLacs, choked jets in core-collapse SN, star burst galaxies, low-luminosity BLLacs, and GRBs.
For each model we perform a model comparison with respect to the single power law, which serves as a benchmark model.
To compute the null hypothesis evidence we marginalize over the parameters of the single power law -- {\it i.e.}\ normalization and spectral index -- and the analysis nuisance parameters.
For the alternative models we marginalize over the nuisance parameters while keeping the astrophysical model fixed at its predicted value.
The Bayes factors for these comparisons are given in the column labeled ``Model only''.
Since some models are not expected to explain the complete astrophysical component, but only parts of the spectrum, we also perform a model comparison where we compare the benchmark single power law against one of the {\it ad hoc}\ models with an additional unbroken power law.
The Bayes factors for these comparisons are given in the column labeled ``Model + SPL''.
Results of these analyses reported in Tbl.~\ref{tbl:diffuse_models} fall into two categories:
\begin{itemize}
    \item Models that are significantly worse compared to the single power law, and for which the addition of a single power-law component does not significantly improve the Bayes factor.
    \item Models that are significantly poorer in comparison to the single power law, but for which the addition of a single power-law component improves its stance compared to the benchmark model.
\end{itemize}

\begin{table*}[h!]
    \centering
    \begin{minipage}{\linewidth}
    \begin{tabular*}{\textwidth}{l r r r r}
        \toprule
        \multirow{2}{*}{\makecell[l]{Model}} & \multirowcell{2}{Model only \\ Bayes factor} & \multirowcell{2}{Model + SPL \\ Bayes factor} & \multirowcell{2}{Most-likely \\ SPL $\astrodeltagamma$} & \multirowcell{2}{Most-likely \\ SPL $\astronorm$} \\
         & & & & \\ \midrule
        \SteckerTableSummary \\ \midrule
        \FangTableSummary \\ \midrule
        \KimuraBOneTableSummary \\ \midrule
        \KimuraBFourTableSummary \\ \midrule
        \KimuraTwoCompTableSummary \\ \midrule
        \MariaBLLacsTableSummary \\ \midrule
        \MurasechockedJetsTableSummary \\ \midrule
        \SBGminBmodelTableSummary \\ \midrule
        \TavecchilowPowerTableSummary \\ \midrule
        \TDEWinterBiehlTableSummary \\ \midrule
        \bottomrule
    \end{tabular*}
    \end{minipage}
    \begin{minipage}{\linewidth}
    \caption{\textbf{\textit{Astrophysical neutrino flux model comparison test results.}} Each row shows the specific model tested, the Bayes factor of the model on its own, the Bayes factor of the model in conjunction with a power-law component, the most likely spectral index of the accompanying power-law component with corresponding $\SigmaOne$ HPD region, and the most likely normalization of the accompanying power-law component with corresponding $\SigmaOne$ HPD region. For the ``Model + SPL'' results the prior factors cancel and so some improper priors are used. For the ``Model only'' results all prior factors except those from the single power-law parameters cancel. In this case we use uniform priors for the astrophysical normalization and astrophysical spectral index that have ranges of $[0,25]$ and $[2,4]$ respectively.} \vspace{-6mm}\label{tbl:diffuse_models}
    \end{minipage}
\end{table*}
Though some scenarios show preference over the null hypothesis, at the moment, none of the performed tests report a Bayes factor greater than ten.
In Jeffreys' scale this corresponds to no strong preference for any of the proposed alternative models~\cite{jeffreys1998theory}.

\section{Conclusion}

We have performed an updated analysis of seven and a half years of the high-energy starting event selection.
For this analysis, the treatment systematic uncertainties has been greatly improved.
The more relevant improvements are: an improved treatment of the atmospheric neutrino background components, a new treatment for Monte Carlo statistical uncertainties, and better characterization of uncertainties in the detector response.
In this analysis, we have studied the observed high-energy astrophysical component by parameterizing it in terms of generic models, such as a single unbroken power law in energy, and also by comparing it to models in the literature.
The best-fit model parameters for the single power law are in agreement with previous iterations of this analysis and are not in tension with results from the combined analysis of IceCube data~\cite{Aartsen:2015knd} or the analysis of northern sky through-going muons~\cite{Aartsen:2016xlq,Haack:2017dxi} at the \SI{95}\percent{} C.L.
Differences between the analysis results may be a symptom of additional spectral structure, but the samples individually do not have power to discern this.
The double power-law component is not preferred with respect to the single power-law hypothesis.
We also report if there is preference over the benchmark single power-law scenario, for models of the astrophysical component available in the literature.
We find that, at the moment, no model is significantly preferred compared to the single power-law description.

\bibliographystyle{ICRC}
\bibliography{references}

\end{document}